\documentclass[conference]{IEEEtran}
\ifCLASSINFOpdf
  \usepackage[pdftex]{graphicx}
  \graphicspath{{../pdf/}{../jpeg/}}
  \DeclareGraphicsExtensions{.pdf,.jpeg,.png}
\else
\fi
\usepackage{hhline}
\usepackage{multirow}
\usepackage{multicol, blindtext}
\usepackage{amsmath}
\usepackage{amssymb}% http://ctan.org/pkg/amssymb
\usepackage{pifont}% http://ctan.org/pkg/pifont

\usepackage{pgfplots}
\pgfplotsset{compat=1.14} 
\usepackage{xcolor,colortbl}
\definecolor{charcoal}{rgb}{0.21, 0.27, 0.31}
\definecolor{chestnut}{rgb}{0.8, 0.36, 0.36}
\definecolor{darkbyzantium}{rgb}{0.36, 0.22, 0.33}
\definecolor{darkorange}{rgb}{1.0, 0.55, 0.0}
\definecolor{darkscarlet}{rgb}{0.34, 0.01, 0.1}
\definecolor{eggplant}{rgb}{0.38, 0.25, 0.32}
\definecolor{frenchrose}{rgb}{0.96, 0.29, 0.54}
\definecolor{battleshipgrey}{rgb}{0.52, 0.52, 0.51}
\definecolor{tiffanyblue}{rgb}{0.04, 0.73, 0.71}
\definecolor{skyblue}{rgb}{0.53, 0.81, 0.92}
\usepackage{amsmath}
\mathchardef\mhyphen="2D % Define a "math hyphen"

\usepackage[normalem]{ulem}
\usepackage{array}
\newcommand{\PLH}{{\mkern-2mu\times\mkern-2mu}}

\usepackage{xcolor}
\usepackage[linesnumbered,ruled,vlined]{algorithm2e}

\SetCommentSty{mycommfont}
\SetKwInput{KwInput}{Input}                % Set the Input
\SetKwInput{KwOutput}{Output}              % set the Output
\SetKwInput{KwWeights}{Weights}              % set the Output
\SetKwInput{KwBias}{Bias}              % set the Output
\SetKw{KwBy}{by}

\usepackage{siunitx}

\usepackage{amsmath}
\usepackage{algorithmicx}
\usepackage[caption=false,font=normalsize,labelfont=sf,textfont=sf]{subfig}
\usepackage{url}
\begin{document}
%
% paper title
% Titles are generally capitalized except for words such as a, an, and, as,
% at, but, by, for, in, nor, of, on, or, the, to and up, which are usually
% not capitalized unless they are the first or last word of the title.
% Linebreaks \\ can be used within to get better formatting as desired.
% Do not put math or special symbols in the title.
\title{MajorityNets: BNNs Utilising Approximate Popcount for Improved Efficiency}

% author names and affiliations
% use a multiple column layout for up to three different
% affiliations
\author{\IEEEauthorblockN{Seyedramin Rasoulinezhad\IEEEauthorrefmark{1}, Sean Fox\IEEEauthorrefmark{1}, Hao Zhou\IEEEauthorrefmark{2}, Lingli Wang\IEEEauthorrefmark{2}, David Boland\IEEEauthorrefmark{1}, Philip H.W. Leong\IEEEauthorrefmark{1}}
\IEEEauthorblockA{\IEEEauthorrefmark{1}School of Electrical and Information Engineering, The University of Sydney, Australia 2006\\
\IEEEauthorrefmark{2}State Key Lab of ASIC and System,
Fudan University, Shanghai 201203, China \\
Email:\{seyedramin.rasoulinezhad, sean.fox, philip.leong, david.boland\}@sydney.edu.au, \{zhouhao, llwang\}@fudan.edu.cn}
}

\iffalse
\author{
\IEEEauthorblockN{The author's names are removed for the peer review process}
\IEEEauthorblockA{}
}
\fi

% for over three affiliations, or if they all won't fit within the width
% of the page, use this alternative format:
% 
%\author{\IEEEauthorblockN{Michael Shell\IEEEauthorrefmark{1},
%Homer Simpson\IEEEauthorrefmark{2},
%James Kirk\IEEEauthorrefmark{3}, 
%Montgomery Scott\IEEEauthorrefmark{3} and
%Eldon Tyrell\IEEEauthorrefmark{4}}
%\IEEEauthorblockA{\IEEEauthorrefmark{1}School of Electrical and Computer Engineering\\
%Georgia Institute of Technology,
%Atlanta, Georgia 30332--0250\\ Email: see http://www.michaelshell.org/contact.html}
%\IEEEauthorblockA{\IEEEauthorrefmark{2}Twentieth Century Fox, Springfield, USA\\
%Email: homer@thesimpsons.com}
%\IEEEauthorblockA{\IEEEauthorrefmark{3}Starfleet Academy, San Francisco, California 96678-2391\\
%Telephone: (800) 555--1212, Fax: (888) 555--1212}
%\IEEEauthorblockA{\IEEEauthorrefmark{4}Tyrell Inc., 123 Replicant Street, Los Angeles, California 90210--4321}}

% use for special paper notices
%\IEEEspecialpapernotice{(Invited Paper)}

% make the title area
\maketitle

% As a general rule, do not put math, special symbols or citations
% in the abstract
\begin{abstract}
Binarized neural networks (BNNs) have shown exciting potential for utilising neural networks in embedded implementations where area, energy and latency constraints are paramount. With BNNs, multiply-accumulate (MAC) operations can be simplified to XnorPopcount operations, leading to massive reductions in both memory and computation resources. Furthermore, multiple efficient implementations of BNNs have been reported on field-programmable gate array (FPGA) implementations. %Since, the XnorPopcount circuits contribute majorly to resource utilisation, the necessity of optimization is raised.
This paper proposes a smaller, faster, more energy-efficient approximate replacement for the XnorPopcount operation, called XNorMaj, inspired by state-of-the-art FPGA look-up table schemes which benefit FPGA implementations. We show that XNorMaj is up to 2$\times$ more resource-efficient than the XnorPopcount operation. While the XNorMaj operation has a minor detrimental impact on accuracy, the resource savings enable us to use larger networks to recover the loss.%accuracy. We demonstrate this by translating the CNV network with XnorPopcount to a padded CNV network with XNorMaj, enhancing accuracy by 2\% for the same FPGA resource cost.

\end{abstract}
% no keywords

% For peerreview papers, this IEEEtran command inserts a page break and
% creates the second title. It will be ignored for other modes.
\IEEEpeerreviewmaketitle
\section{Introduction}
% no \IEEEPARstart

%Recent research on deep neural networks (DNNs) has yielded a significant improvement over other techniques in cognitive domains such as artificial intelligence and robotics~\cite{DBLP:journals/spm/ArulkumaranDBB17}. Among DNNs, convolutional neural networks (CNNs) have achieved high-performance on video and image processing tasks~\cite{DBLP:journals/corr/abs-1905-11946}. The ability of CNNs to achieve such results requires a massive number of parameters and complex computations. This makes them challenging to deploy in real-time.

Recent research on convolutional neural networks (CNNs) has yielded a significant improvement over other techniques in cognitive domains. This ability requires a massive number of parameters and complex computations, which makes them challenging to deploy in real-time. %There has been considerable interest in efficient DNNs for embedded and mobile applications, where common concerns are low-latency and energy while delivering acceptable accuracy. 

Quantization techniques bring significant performance enhancement by reducing both memory footprint and resource requirements of compute units. Binarized neural networks (BNNs) are the most extreme case of quantization, using a single bit for each activation and weight so the majority of energy-hungry multiply~accumulate (MAC) computations can be replaced by the XnorPopcount operation~\cite{DBLP:journals/corr/CourbariauxB16}. %Table~\ref{table_BNN_overview} presents an overview of common BNN architectures and the percentage usage of XnorPopcount. 
Previous research has suggested modifications to field-programmable gate arrays (FPGAs) lookup-table (LUT) structures to enhance the efficiency of popcount operation~\cite{ramin:JHAndersonALMIdea}. Wang~et~al.~\cite{DBLP:conf/fccm/WangDCC19} exploited the capabilities of FPGA LUTs to manipulate XnorPopcount operation to save resources in an fully-unrolled manner. 

To further improve the efficiency of FPGA-based BNN architepoolingctures, we propose a smaller and faster approximation for XnorPopcount. We call this XNorMaj, based on integrating Majority and popcount circuits. We report on its efficiency on FPGA platforms, and show that it offers significant reductions in area and critical path. This is the first work focused on simplifying XnorPopcount operations in a fold-able manner with consideration of FPGA architectures. In summary, the contributions of this work:

\begin{itemize}
  \item A novel approximate replacement for XnorPopcount, called XNorMaj, leading to new BNN architectures, MajorityNets, using the proposed XNorMaj operation.
  %\item A novel approximate replacement for XnorPopcount, called XNorMaj
  %\item A new DNN architecture, MajorityNets, using the proposed XNorMaj operation.
  \item A quantitative evaluation of impacts of the above techniques on various BNN architectures, considering different performance metrics such as accuracy, area, and delay.
\end{itemize}

\begin{figure}[!t]
\centering
\includegraphics[width=\linewidth]{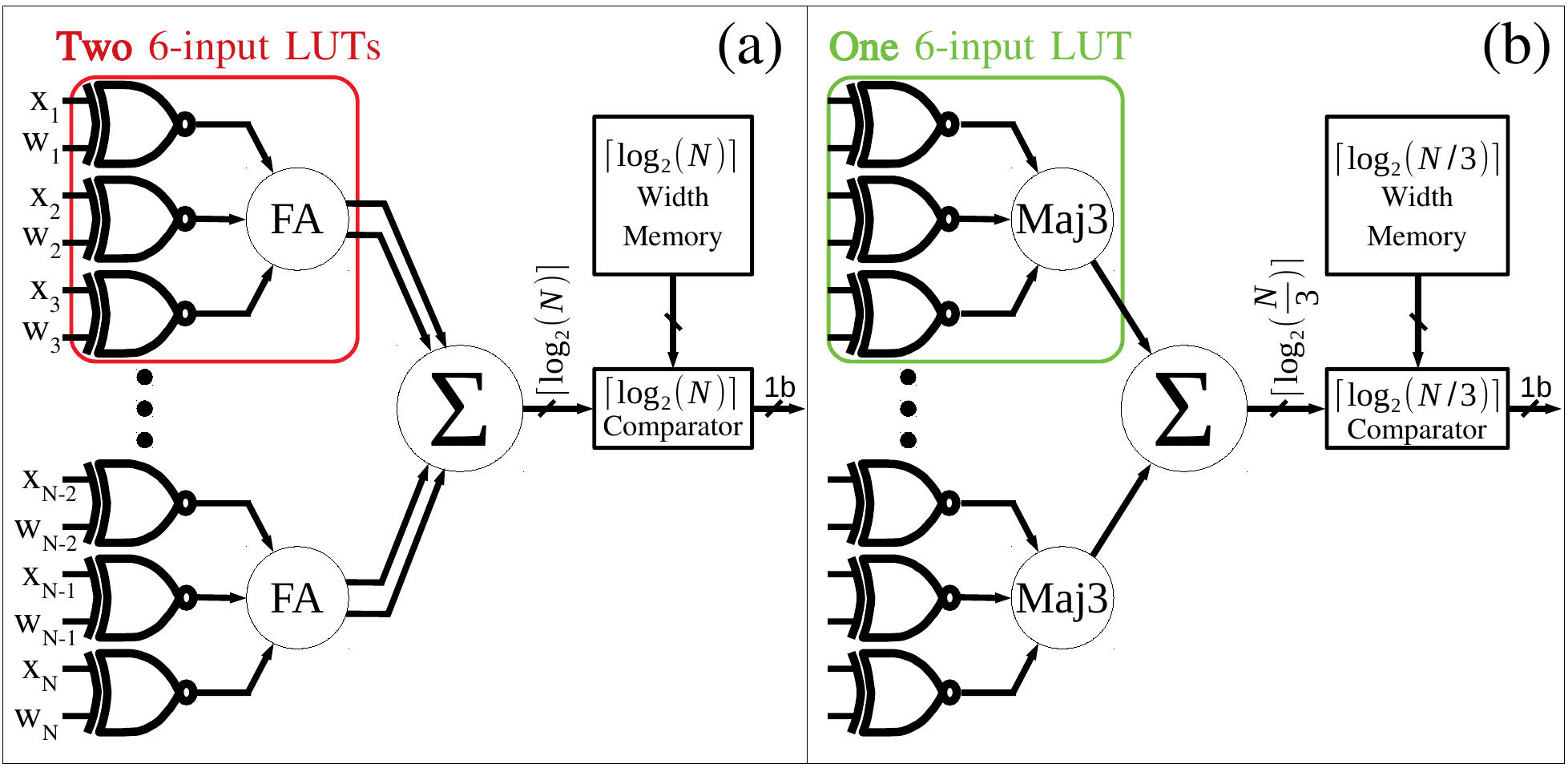}
\caption{a) an XnorPopcount operation and b) an XNorMaj-3 operation and their following threshold layer (according to~\cite{DBLP:conf/fpga/UmurogluFGBLJV17})}
\label{fig:XnorPopcount}
\end{figure}

Verilog models together with a training platform are available as open source software on github.com/raminrasoulinezhad/MajorityNets.

\section{XNorMaj-$M$ Popcount (XNorMaj) operation}
%This section shows how majority circuits can approximate the XnorPopcount operation. First, XnorPopcount operation is introduced. Second, based on using majority circuits, our new XNorMaj operation is suggested, and its superiority for FPGA platforms are described. Then, using the new operation, the majority convolution (MConv) and fully connected (MFC) layers are proposed.  

\subsection{XnorPopcount operation}
In CNNs, convolution (Conv) and fully-connected (FC) layers comprise the majority of computations. Each single output of both layers can be modeled by a neuron-like computation as $y = \sum^{N-1}_{i=0}x_{i}w_{i} + B$.
\iffalse
\begin{equation}
\begin{split}
y & = \sum^{N-1}_{i=0}x_{i}w_{i} + B
\label{eq:neurun_normal}
\end{split}
\end{equation}
\fi
By constraining the activations and weights to +1 and -1, and representing them by logic 1 and 0 respectively, multiplication of each pair can be done using an XNOR gate, as proposed by Courbariaux~et~al.~\cite{DBLP:journals/corr/CourbariauxB16}. Thus, high-precision MAC operations can be simplified to bit-wise XNOR operations followed by a counter, called XnorPopcount~\eqref{eq:neurun_binary}.
\begin{equation}
\begin{split}
    %y & = \sum^{N-1}_{i=0}x^{b}_{i}w^{b}_{i} + b \\
    %  & = 2\times \sum^{N-1}_{i=0}Xnor(x^{b}_{i},w^{b}_{i}) - N + b
      %& = 2\times XnorPopcount(\mathbf{\hat{x}},\mathbf{\hat{w}}) - N + b
      y = 2\times \sum^{N-1}_{i=0}Xnor(x^{b}_{i},w^{b}_{i}) - N + B
\label{eq:neurun_binary}
\end{split}
\end{equation}

%CPUs and GPUs support bit-wise XNOR and popcount through specialized instructions~\cite{DBLP:conf/vlsid/RamanarayananMEKG08}. In ASIC implementations, XNOR gates followed by a compressor tree can be used to build the XnorPopcount circuits. In FPGA implementations, XnorPopcount circuit is mapped to LUTs and ternary-adders {\em e.g.}~\cite{DBLP:journals/ijon/LiangYLLW18}. 
By implementing XnorPopcount operation on different FPGA architectures, we observed that the implementation of Xnor gates and primary compressor circuits are fused. As depicted in Figure~\ref{fig:XnorPopcount}(a), every three couples of activations and weights are assigned to two LUTs to implement the three XNOR gates and their following Full-Adder (FA) counter. Then, the 2-bit answers of the mentioned blocks are summed by a compressor tree. According to this structure, the first two LUTs offer a compression rate of 3 input pairs:2. %Also, since in practice, the number of inputs for a neuron in Conv and FC layers is in the thousands. 
%Also, as Table~\ref{table_NormalPopcountResources} shows, for large input sizes, resource requirements and speed are the main concerns. 

\iffalse
\begin{table}
\renewcommand{\arraystretch}{1.3}
\setlength{\tabcolsep}{4.5pt}
\caption{The resource usage of XNOR, Popount, and XnorPopcount operations, Delay (ns), (using Vivado2018.2)}
\label{table_NormalPopcountResources}
\centering
\begin{tabular}{|l||c||c|c||c|c|}
    \hline
    \multirow{2}{*}{Input size} & {XNOR} & \multicolumn{2}{c||}{Popcount} & \multicolumn{2}{c|}{XnorPopcount} \\
    \cline{2-6}
    {} & {LEs} & {LEs} & {Delay} & {LEs} & {Delay}\\
    \hline
    \hline
    {$3\PLH3\PLH64$} & {576}& {750} & {4.36} & {849} & {4.50}\\
    \hline 
    {$3\PLH3\PLH128$} & {1152}& {1704} & {4.92} & {1880} & {5.22}\\
    \hline 
    {$3\PLH3\PLH256$} & {2304} & {3546} & {5.86} & {3820} & {5.84}\\
    \hline 
    {$3\PLH3\PLH512$} & {4608} & {7219} & {5.98} & {7843} & {6.00}\\
    \hline 
\end{tabular}
\end{table}
\fi

\subsection{XNorMaj technique}

To achieve further compression, we first rewrite Equation~\eqref{eq:neurun_binary} using a two-level hierarchical summation (assume ${\hat{Z}} = \textrm{XNOR}(\mathbf{\hat{x}},\mathbf{\hat{w}})$),
\begin{equation}
\begin{split}
y = 2 \sum^{\frac{N}{M}-1}_{i=0}\sum^{M-1}_{m=0}\hat{Z}_{iM+m} - N + B
\label{eq:neurun_rewrite}
\end{split}
\end{equation}

We then approximate the inner loop in Equation~\eqref{eq:neurun_rewrite} with a scaled $M$-input Majority circuit (Maj-$M$), which indicates whether more than half of the inputs are $True$. %Altogether, this is described by~\eqref{eq:neurun_proposed1}. 
This new operation, called XNorMaj, involves a bit-wised XNOR applied between the input activations and their corresponding weights, with each $M$-grouped output passed to a Maj-$M$ circuit to generate a single bit result which is scaled by $V^{1}_{i}$ and $V^{0}_{i}$.

\begin{equation}
\begin{split}
\widetilde{y}\!&= 2\!\!\sum^{\frac{N}{M}-1}_{i=0}\!\!(\mathbb{M}^{M-1}_{m=0}(\hat{Z}_{iM+m}) \PLH (V^{1}_{i}\!\!-\!V^{0}_{i})\!+\!V^{0}_{i}) \!-\!N\!\!+\!B
\label{eq:neurun_proposed1}
\end{split}
\end{equation}
where Maj-$M$ = $\mathbb{M}^{M-1}_{m=0}=\left\{
  \begin{array}{@{}ll@{}}
    1, & \text{if } \sum_{i=0}^{M-1}{x_{i}}\geq M/2 \\
    0, & \text{otherwise}
  \end{array}\right.$.
\iffalse
\begin{align}
%\begin{split}
\widetilde{y}\!= 2&\!\!\sum^{\frac{N}{M}-1}_{i=0}\!\!(\mathbb{M}^{M-1}_{m=0}(\hat{Z}_{iM+m}) \PLH (V^{1}_{i}\!\!-\!V^{0}_{i})\!+\!V^{0}_{i}) \!-\!N\!\!+\!B \label{eq:neurun_proposed1}\\
\textrm{where }  & \textrm{Maj-$M$}=\mathbb{M}^{M-1}_{m=0}=
  \begin{cases}
    1,& \text{if } \sum_{i=1}^{M}{x_{i}}\geq M/2\\
    0,& \text{otherwise}
\end{cases}
\label{eq:maj_M}
\end{align}
\fi 
By using a common scaling factor this simplifies to popcounting the single bit outputs of the majority circuits and scaling appropriately (Equation~\eqref{eq:neurun_proposed_final}).

\begin{align}
\widetilde{y}\!&=\!\!2(V^{1}\!-\!V^{0})\!\!\sum^{\frac{N}{M}-1}_{i=0} \!\!\mathbb{M}^{M-1}_{m=0}(\hat{Z}_{iM+m})\!+\! \frac{N\PLH V^{0}}{M}\!-\!N\!\!+\!\!B \label{eq:neurun_proposed_final}%\\
%\end{split}
\end{align}

% new version
In Equation~\eqref{eq:neurun_proposed_final}, the majority circuit scale factors and bias value implement a linear transform (LT) of the neuron output. The transform is similar for all neurons in the same channel. The computation of a batch normalization (BN) layer is also a LT modeled as $\widehat{y}=\gamma(\widetilde{y}-\mu)i+\beta$%:
\iffalse
\begin{equation}
\begin{split}
\widehat{y} &= \gamma(\widetilde{y} - \mu )i+\beta
\label{eq:BN}
\end{split}
\end{equation}
\fi
, where ($\gamma$, $\mu$, $i$ and $\beta$) are the BN layer parameters defined per channel~\cite{DBLP:conf/icml/IoffeS15}. In cases where Conv or FC layers are followed by a BN layer, these two LT functions can be merged into a single LT function. In practice, a BN layer without separate scaling layer is sufficient and this forces BN parameters to adapt scaling factors for majority circuits in each output channel separately. Furthermore, by following the assumption in~\cite{DBLP:conf/fpga/UmurogluFGBLJV17}, the new LT function can be merged with activation function in a threshold layer. Because of the mentioned reasons, we fixed the $V^{1}$ and $V^{0}$ values to be 2.625 and 0.375 respectively, and used a BN layer after each layer using XNorMaj operation.

Using the Majority circuit introduces new trade-offs. Since majority circuits compress $M$-grouped inputs into a single bit, the popcount is reduced by a factor of $M$, leading to a smaller popcounter circuit. The comparator circuits and threshold parameters are also simplified and hence smaller. Unfortunately, the technique results in reduced accuracy. The accuracy-performance trade-offs are therefore dependent on the choice of parameter $M$; for this paper, we focus on XNorMaj-3 ($M=3$) operations for the following reasons:
\begin{itemize}
  \item {\bf Implementation efficiency for FPGA platforms:} As demonstrated in Figure~\ref{fig:XnorPopcount}.b, three XNOR gates, and the following Maj-3 circuits can be combined and mapped to a single 6-input LUT. This fused computation is fed by the same input scheme comparing to the baseline implementation of three XNOR gates and the following FA, XNorFA (Figure~\ref{fig:XnorPopcount}.a). However, it produces a one-bit output rather than two bits leading to smaller compression trees. By synthesizing different XNorMaj-$M$ circuits for FPGAs in Table~\ref{table:XnorMaj3Popcount_proof}, it can be seen that XNorMaj-3 unite offers the best compression rate vs. complexity trade-off. 

  \item {\bf Accuracy:} by increasing the parameter $M$, the similarity of the XNorMaj-$M$ popcount and the baseline model is reduced, and inference accuracy drops.

  \item {\bf Integration with Conv layer kernels:} The number of input pairs for a neuron in an convolution layer is multiple of kernel spatial dimensions. Since $M$ has to be an odd number, by choosing $M$ equal to the kernel size, which is also an odd number, and applying majority logic on the pairs placed in the same channel in a row or a column, folding is possible. Also, three is the most common kernel size for Conv layers in modern BNNs~\cite{DBLP:journals/ijon/LiangYLLW18}.
  
\end{itemize}

\begin{table}
\renewcommand{\arraystretch}{1.3}
\setlength{\tabcolsep}{1.75pt}
\caption{FPGA implementation efficiency of XNorMaj-$M$ unite. Reported delays measured by registering inputs and outputs. Eff: Compression rate /(LUTs $\PLH$ Delay), Compression rate = (number of input pairs):(output width), Delay: $ns$}
\label{table:XnorMaj3Popcount_proof}
\centering
\begin{tabular}{|c||c|c|c|c|c|c|}
    \hline
    \multirow{2}{*}{Device} & {Metrics} & {XNorFA} & {XNorMaj-3} & {XnorMaj-5} & {XnorMaj-7} & {XnorMaj-9}\\
    \cline{2-7}
    {} & {Comp.} & {3P:2} & {3P:1} & {5P:1} & {7P:1} & {9P:1}\\
    \hline 
    \hline 
    \multirow{3}{*}{Xilinx} & {LUT} & {2} & {\bf 1} & {3} & {5} & {7}\\
    \cline{2-7}
    {} & {Delay} & {0.68} & {\bf 0.64} & {1.10} & {0.99} & {1.07}\\
    \cline{2-7} 
    {} & {Eff.} & {1.11} & {\bf 4.67} & {1.52} & {1.41} & {1.20}\\
    \hline 
    \hline 
    \multirow{3}{*}{Intel} & {ALMs} & {2} & {\bf 1} & {3} & {5} & {9}\\
    \cline{2-7}
    {} & {Delay} & {0.86} & {\bf 0.70} & {0.96} & {1.24} & {1.78}\\
    \cline{2-7} 
    {} & {Eff.} & {0.87} & {\bf 4.26} & {1.74} & {1.13} & {0.56}\\
    %\hline 
    %\hline 
    %\multirow{2}{*}{ASIC} & {Area} & {65.5} & {\bf 48.96} & {102.24} & {124.56} & {203.04}\\
    %\cline{2-7}
    %\multirow{2}{*}{65-nm} & {Delay} & {0.38} & {\bf 0.30} & {0.39} & {0.44} & {0.51}\\
    %\cline{2-7} 
    %{} & {Power} & {0.32} & {\bf 0.12} & {0.17} & {0.19} & {0.26}\\
    %\cline{2-7} 
    %{} & {Eff.} & {0.06} & {\bf 0.20} & {0.13} & {0.13} & {0.09}\\
    \hline 
\end{tabular}
\end{table}

\subsection{Majority Convolution and Fully-connected layers}
Consider a standard convolutional layer which takes a $D_F \times D_F \times M$ feature map $\mathbf{F}$ as input, and produces a $D_G \times D_G \times N$ feature map $\mathbf{G}$ as output. The output is generated via a convolution with a $D_K \times D_K \times M \times N$ kernel $\mathbf{K}$ and addition with a $N$-element bias vector $\mathbf{B}$. Algorithm~\ref{algorithm:ConvLayer_mixed} describes the computation %when XnorMaj is not enabled. Replacing the popcount operation with Majority popcount, by enabling XNorMaj in Algorithm~\ref{algorithm:ConvLayer_mixed}, Majority convolution (MConv) is proposed.
for standard and majority convolution (MConv) cases.
Since binarization is applied on activations and weights, we modeled the majority circuit and the previously mentioned scaling factors using clip and scale functions. Scaling factor should be selected to be consistent with Equation~\eqref{eq:neurun_proposed_final}. Using the same approach, a Majority fully-connected (MFC) layer can be derived.

\begin{algorithm}[t]
\label{algorithm:ConvLayer_mixed}
\caption{Standard/Majority-$D_{K}$ convolution layer}
\DontPrintSemicolon
    \For{$n\gets0$ \KwTo $N$}{
        \For{$k\gets0$ \KwTo $D_{G}$}{
            \For{$l\gets0$ \KwTo $D_{G}$}{
                \For{$m\gets0$ \KwTo $M$}{
                    \For{$i\gets0$ \KwTo $D_{K}$}{
                        \If{XNorMaj is enabled}{
                            array $t_{in} \gets F_{k+i-1,:,m}$\\
                            array $t_{w}  \gets K_{i,:,m,n}$\\
                            $t_{d}= t_{in} \cdot t_{w}$ \tcp*[f]{Dot product}\\
                            %$t_{d}= K_{i,:,m,n} \cdot F_{k+i-1,:,m}$ \tcp*[f]{Dot product}\\
                            $G_{k,l,n} \mathrel{+}= clip(t_{d}, (\text{-}1,1))\PLH Scale$
                        }
                        \Else{
                            \For{$j\gets0$ \KwTo $D_{K}$}{
                                $t_{in} \gets F_{k+i-1,l+j-1,m}$\\
                                $t_{w}  \gets K_{i,j,m,n}$\\
                                $G_{k,l,n} \mathrel{+}= t_{in} \times t_{w}$
                            }
                        }
                    }
                }
                $G_{k,l,n} \mathrel{+}= B_{n} $
            }
        }
    }
\end{algorithm}

\begin{figure}[!t]
\centering
\includegraphics[width=\linewidth]{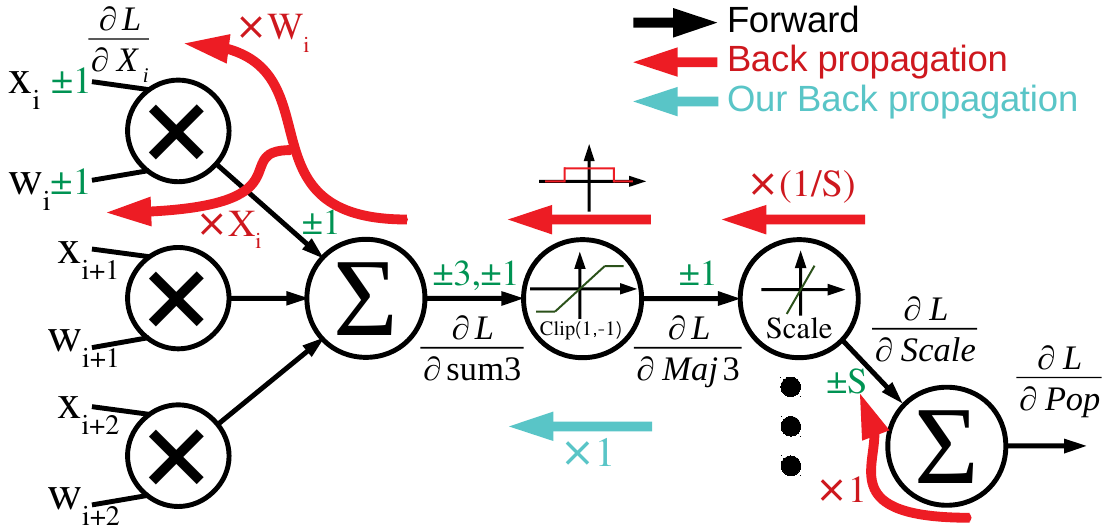}
\caption{Back-propagation computation for Majority layers}
\label{fig:backward}

\end{figure}

The back-propagation algorithm for majority layers can be implemented by applying the chain rule as shown by the red arrows in Figure~\ref{fig:backward}. %However, because of performance issues, the clip and scale functions require different handling. In the case of clipping, a large number of intermediate values are involved, which significantly slows down training. Thus, we use a 
using a
straight through estimator (STE) as a proxy for the derivative of the clipping function (teal arrow)%~\cite{DBLP:journals/corr/abs-1903-05662}
. Moreover, since the scale factor is fixed for a layer, it can be applied directly on computed activation and weight gradients. With these two modifications, Majority layer back-propagation can be simplified as normal layer back-propagation.

%Consider a standard convolutional layer which takes a $D_F \times D_F \times M$ feature map $\mathbf{F}$ as input, and produces a $D_G \times D_G \times N$ feature map $\mathbf{G}$ as output. The output is generated via a convolution with a $D_K \times D_K \times M \times N$ kernel $\mathbf{K}$ and adding with a $N$ bias kernel $\mathbf{B}$ as follows:
%\begin{equation}
%\mathbf{G}_{k,l,n} = \sum_{i,j,m}\mathbf{K}_{i,j,m,n} %\mathbf{F}_{k+i-1,l+j-1,m} + \mathbf{B}_{n}
%\label{eq:convlayer}
%\end{equation}
%By applying binarization on standard convolution as described in \cite{DBLP:journals/corr/CourbariauxB16} the equation can be modified as follows:
%\begin{equation}
%\begin{split}
%\mathbf{G}^{b}_{k,l,n} & = \sum_{i,j,m}\mathbf{K}^{b}_{i,j,m,n} %\mathbf{F}^{b}_{k+i-1,l+j-1,m} + \mathbf{B}_{n}
%\label{eq:convlayer_b}
%\end{split}
%\end{equation}

%%%%%%%%%%%%%%%%%%%%%%%%%%%%%%%%%%%%%%%%%%%%%%%%%%%%%%%%%%%%%%%%%%%%
\section{Results}
%This section shows the hardware efficiency and accuracy trade-offs of XNorMaj operations in BNN implementations.

\subsection{Hardware efficiency of XNorMaj vs. XnorPopcount}
\begin{figure}[!t]
\centering
\includegraphics[width=\linewidth]{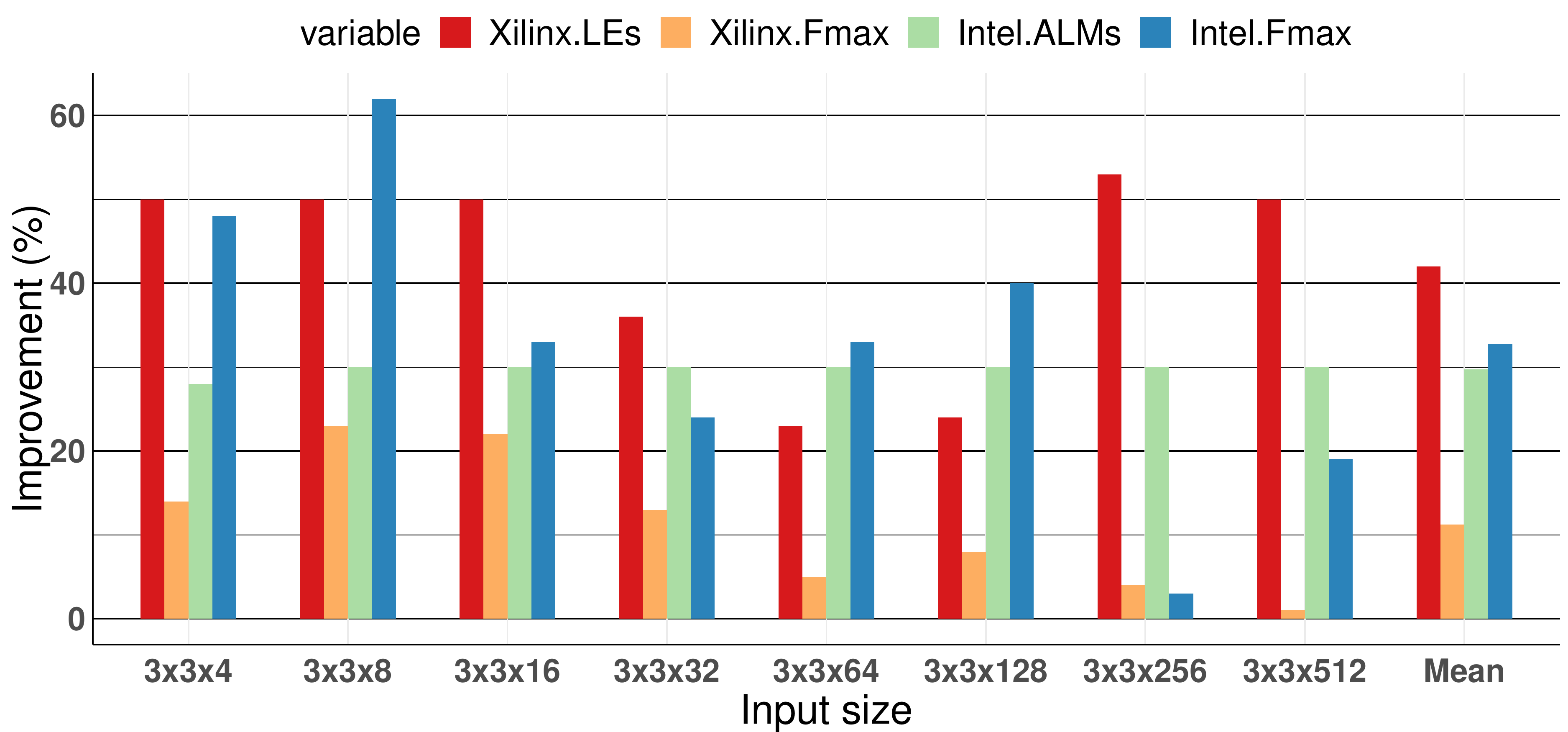}
\caption{XNorMaj-3 improvements using Xilinx/Intel FPGA architectures (Zynq Ultrascale{+} / Aria~10). $F_{max}$: Maximum working frequency}
\label{fig_FPGA}
\end{figure}

We synthesized, placed, and routed a Verilog model of XnorPopcount and XNorMaj circuits using several input sizes for Aria-10 (10AX016E4F29M3SG) and Zynq UltraScale+ (xczu3eg-sbva484-1-e) using Quartus-II 2017.0 and Vivado 2018.2 respectively. As Figure~\ref{fig_FPGA} shows, XNorMaj is 20-50\% smaller, especially for the large input sizes. We also observed that critical paths are dramatically reduced in Intel architectures, with a smaller gain on Xilinx FPGAs. %We believe the gap between the two architectures is because of differences between LUT architectures. The reduction delay is the result of the better compression rate of XnorMaj-3 comparing to the XnorFA unite, which leads to a fewer number of logic layers in the compression tree phase. Similar results are archived for Stratix-10 and Virtex7 UltraScale architectures.

\subsection{Hardware Efficiency of MConv and MFC layers}
\begin{figure}[!t]
\centering
\includegraphics[width=\linewidth]{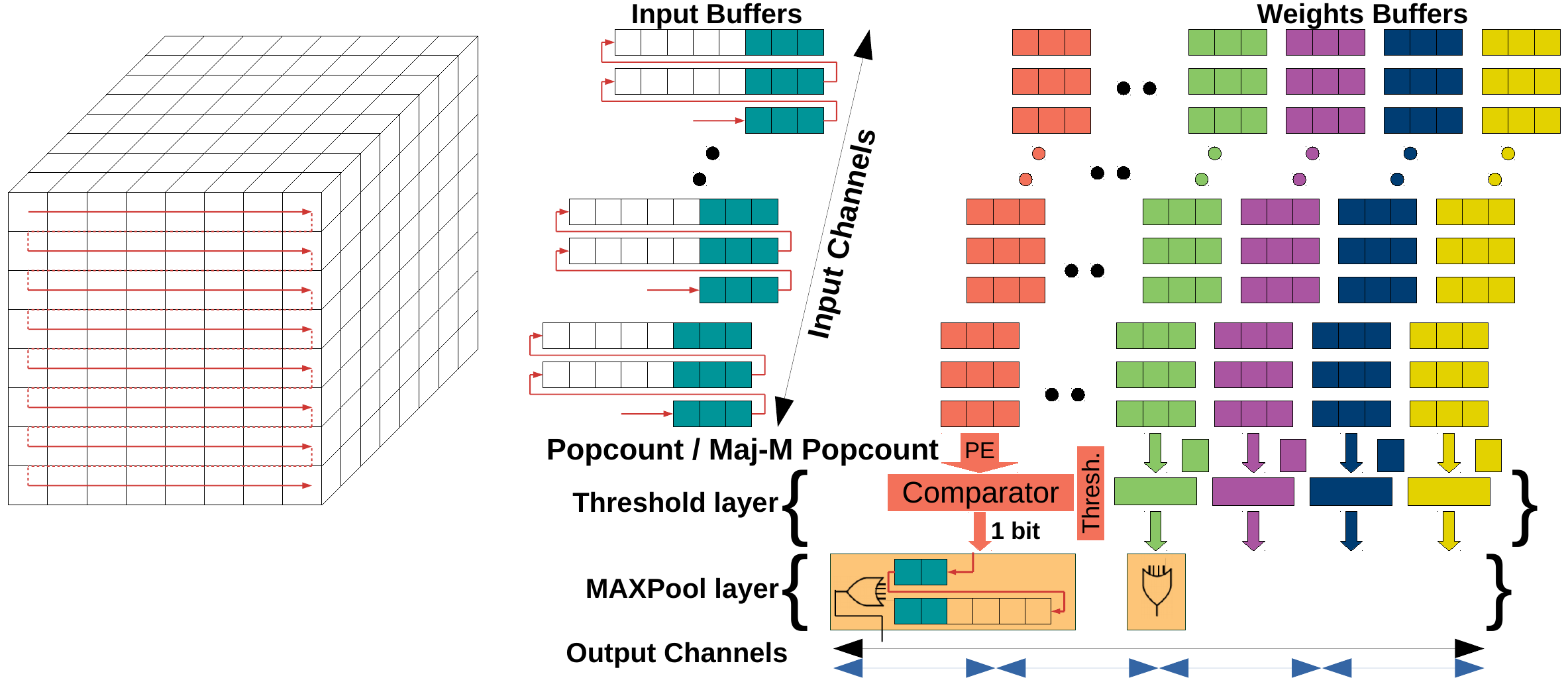}
\caption{Illustration of accelerator operation}
\label{fig_unrooled_accelerator}
\end{figure}

Figure~\ref{fig_unrooled_accelerator} demonstrates the Conv/FC layer implementation used to measure the efficiency of XNorMaj-3. In this implementation, the input feature map is streamed over channels in parallel. Each stream is saved in a channel buffer to provide a window of $D_K \times D_K \times M$ activations to all processing units (PU). Each PU is responsible for computing pixels of a channel of the output feature map. For FC layers, we assume inputs are available in parallel and there is no need for sliding. If Conv/FC layers are followed by a pooling layer, with the same buffering scheme, the pooling layer is implemented on top of that layer. Also, a layer following pooling layers can be folded to keep the same throughput rate, achieving full utilization. For instance, a Conv layer after a Maxpool layer with a $2\times 2$ kernel size should be folded 4$\times$ more than the previous layer. In our design, we fold the number of PEs rather than their input size by folding factor ($FF$). This approach prevents high-precision MAC operations on partial results. 

Table~\ref{table:MCov_FPGA} summarises the implementation results for different layers of the padded version of CNV~\cite{DBLP:conf/fpga/UmurogluFGBLJV17} (CNV-P) network, where all Conv layers are using padding. Using XNorMaj-3 reduces the required LUTs by 20-43\% per layer and 30\% in total. By resynthesizing the layers regardless of throughput balancing, the logic element (LE) reduction grows to 43\% which shows the affect of folding on reduction rate (last column of the Table~\ref{table:MCov_FPGA}). To measure the efficiency for highly-folded implementations we increased the $FF$ by 8$\times$ for all layers which limits the resource reduction to 13\%.

\begin{table}
\renewcommand{\arraystretch}{1.3}
\setlength{\tabcolsep}{3pt}
%\vspace{-1ex}
\caption{LE usage comparison of (M)Conv and (M)FC layers using CNV-P. (layer \#1 is not included), $^{\star}$: padded to be multiple of 3}
\label{table:MCov_FPGA}
\centering
\begin{tabular}{|c|c|c|c||c|c||c|c|}
    \hline
    \multicolumn{4}{|c||}{Layer configurations} & {Conv} & {MConv} & \multicolumn{2}{c|}{LE Improvement}\\
    \cline{1-4}
    \cline{7-8}
    {\#} & {Cin} & {Cout} & {$FF$} & {/FC} & {/MFC} & {folded} & {non-folded}\\
    \hline
    \hline
    {Conv2} & {64} & {64} & {1} & {55k} & {40k} & {27\%} & {27\%}\\
    \hline 
    {Conv3} & {64} & {128} & {4} & {38k} & {30k} & {20\%} & {26\%}\\
    \hline 
    {Conv4} & {128} & {128} & {4} & {86k} & {63k} & {27\%} & {36\%}\\
    \hline 
    {Conv5} & {128} & {256} & {16} & {43k} & {33k} & {27\%} & {36\%}\\
    \hline 
    {Conv6} & {256} & {256} & {16} & {87k} & {50k} & {43\%} & {56\%}\\
    \hline 
    \hline
    {FC1} & {4096$^{\star}$} & {512} & {64} & {96k} & {67k} & {30\%} & {36\%}\\
    \hline 
    {FC2} & {512$^{\star}$} & {512} & {64} & {11k} & {8k} & {22\%} & {36\%}\\
    \hline 
    {FC3} & {512$^{\star}$} & {10} & {10} & {1K} & {0.7k} & {29\%} & {30\%}\\
    \hline 
    \hline 
    \multicolumn{4}{|c||}{Total} & {417k} & {291k} & {30\%}&{43\%}\\
    \hline  
\end{tabular}
\end{table}

\subsection{Accuracy}
To explore the effect of using XNorMaj on the BNNs, we trained different models on different datasets. The training platform is available on the GitHub repository, which is based on the open-source project in reference~\cite{DBLP:conf/nips/HubaraCSEB16}.

First, we applied the proposed idea on two multi-layer perceptrons, SFC and LFC networks~\cite{DBLP:conf/fpga/UmurogluFGBLJV17}. By replacing their all three FC layers with MFC layers, the accuracy drop for the MNIST dataset is 0.2\% and 0.3\% while reducing the LEs by 45\% and 35\% respectively for SFC and LFC in the mentioned implementation method with no folding. In the same approach, using CNV-P~\cite{DBLP:conf/fpga/UmurogluFGBLJV17} and VGG-like~\cite{DBLP:conf/nips/HubaraCSEB16} networks, by replacing 2nd-6th Conv layers with MConv, the LE reduction is about 23\% and 30\% with the cost of 1.7\% and 0.5\% accuracy drop for CNV-P and VGG-like respectively (Table~\ref{table:Acc}).

The area reduction makes the CNV-P with majority layers implementation the same cost as a non-padded with XnorPopcount operation. However, by using padding, in CNV-P, convolution layers receive and produce larger feature maps which increase the computations and keep the weight parameters the same for those layers. It also affects the first FC layer, where the number of inputs is increased, {\em e.g.}, 16 times in the case of in the first MFC layer of CNV-P network leading to 16 times more computation and parameters. Since the error rate of adding padding and using majority layers is 2\% less, we recovered accuracy using the same area. 

%In another experiment, for ResNet-20, we replace first convolution layer of each residual blocks with MConv layers. As we measured the accuracy drop for CIFAR10 and CIFAR100 is about 1\% and 1.5\% respectively.  

\begin{table}
\renewcommand{\arraystretch}{1.3}
\vspace{-1.5ex}
\setlength{\tabcolsep}{3pt}
\caption{Error rate (\%) of different BNNs}
\label{table:Acc}
\centering
\begin{tabular}{|l||l|c|c|c|c|c|}
    \hline
    \multirow{2}{*}{BNNs} & \multirow{2}{*}{Dataset} & \multicolumn{2}{c|}{Error(\%) } & \multicolumn{2}{c|}{details} &{LE} \\
    \cline{3-6}
    {} & {} & {baseline} & {Ours} & {MConv} & {MFC} & {Improve}\\
    \hline\hline
    {SFC} & {MNIST} & {3.47\%} & {3.75\%} & {-} & {All} & {45\%}\\
    \hline
    {LFC} & {MNIST} & {2.66\%} & {2.86\%} & {-} & {All} & {35\%}\\
    \hline
    {CNV-P} & {SVHN} & {5.67\%} & {6.28\%} & \multirow{2}{*}{except 1st} & {-} & \multirow{2}{*}{23\%}\\
    \cline{1-4}
    \cline{6-6}
    {CNV-P} & {CIFAR10} & {13.35\%} & {15.01\%} & \multirow{2}{*}{Conv} & {-} & {}\\
    \cline{1-4}
    \cline{6-7}
    {VGG-like} & {CIFAR10} & {10.78\%} & {11.24\%} & {} & {-} & {30\%}\\
    %\hline
    %{ResNet-20} & {CIFAR10} & {11.42\%} & {12.41\%} & {1st half of} & {-} & {-}\\
    %\cline{1-4}
    %\cline{6-7}
    %{ResNet-20} & {CIFAR100} & {\color{red}42.14\%} & {\color{red}39.54\%} & {ResBlocks} & {-} & {-}\\
    \hline
\end{tabular}
\end{table}

\begin{figure}[!t]
\centering
\includegraphics[width=\linewidth]{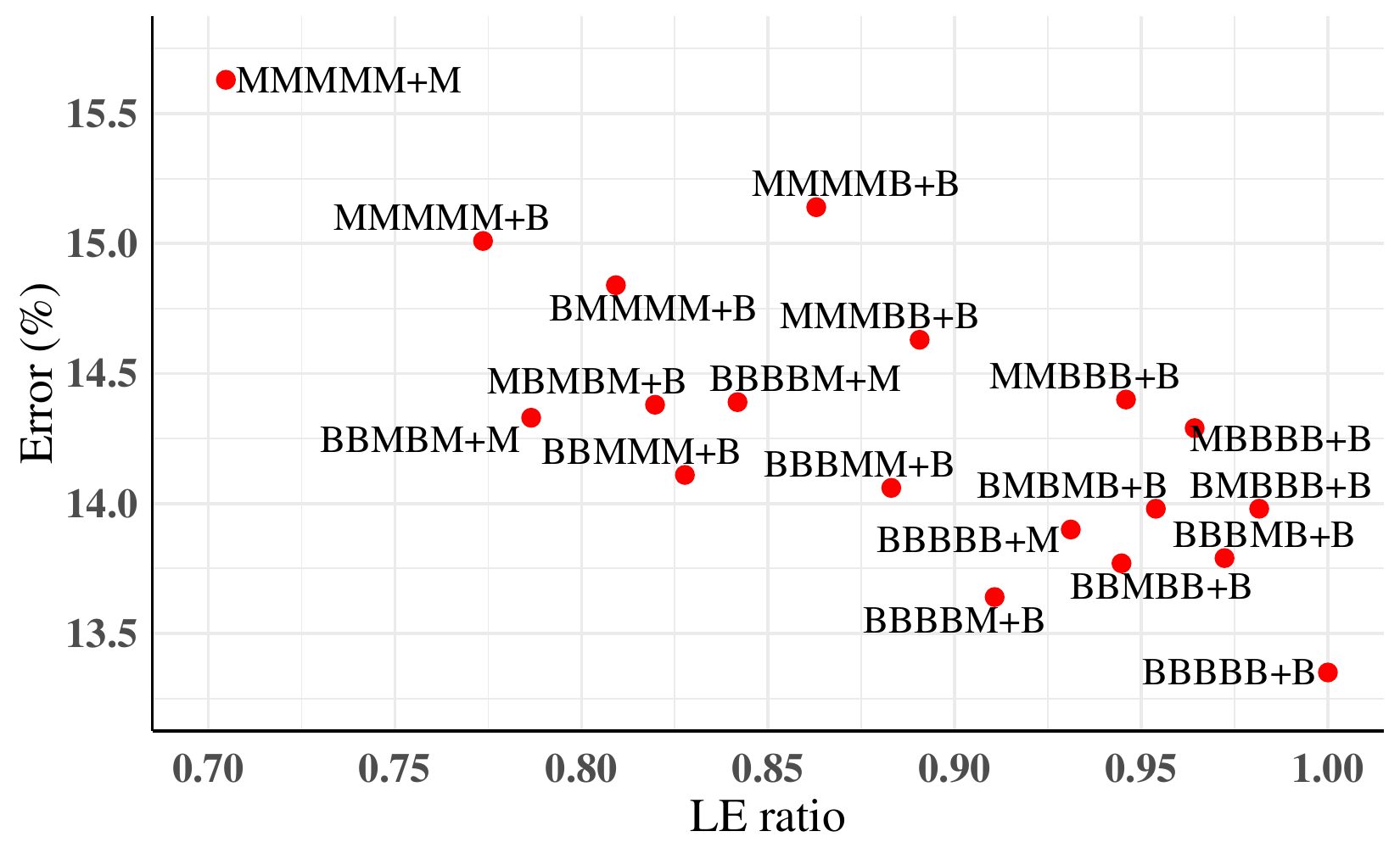}
\caption{Comparison of accuracy vs. LE usage (points on the lower left represent a Pareto set of efficient implementations)}
\vspace{-2ex}
\label{fig:CNV_padded_XXXXX}
\end{figure}

In addition, we explored the effect of arbitrary picking the layers for deployment of XNorMaj-3 operation. We select CNV-P network and the most contributor layers in terms of LEs which are Conv2-6 and the first FC layers. We use a six-character notation, where each character can be B or M, representing whether a layer is using XnorPopcount or XNorMaj. The first five characters specify the configuration of Conv2-6 layers and the last character gives the FC1 layer with ``+" being used to separate the last character. %For instance, MBBMB+M means that the Conv2 and Conv5 and the first fully-connected layers are using XNorMaj while other layers are using XnorPopcount. 
Figure~\ref{fig:CNV_padded_XXXXX} shows performance per LUT of different configurations using our mentioned architecture, enabling a user to trade accuracy for performance by following the Pareto-optimal curve. As an example, the BBMBM+M configuration delivers saves 22\% of LEs with  accuracy reduced by only 1\%.

%As we observed:
%\begin{itemize}
%    \item Replacing the last convolution layers generally offers a better accuracy vs. cost trade-off. {\em e.g.}, BBBBM+B is better than MBBBB+B.  
%    \item Since 4th and 6th Conv and the first FC layers contribute most of the implementation costs, replacing them gives the best accuracy-cost trade-off. Training the model using BBMBM+M configuration shows 22\% LUT reduction in the cost of 1\% accuracy drop.
%\end{itemize}

\section{Conclusion}
We proposed XNorMaj-$M$ popcount, a new approximate XnorPopcount operation, that reduces resource requirements. Compared to a conventional implementation using XNOR and a compression tree, XNorMaj is 20-50\% more area efficient in FPGA platforms. Furthermore, the technique enjoys an average  20\% critical path reduction for Xilinx and Intel FPGA architectures. Using XNorMaj, an semi-unrolled, padded version of the CNV network with the same LUT utilization enjoys 2\% better accuracy. 
In future work we will explore the effect of using Maj-5/7/9 circuits. In addition to that, partially usage of the XNorMaj in a layer would be explored. We will also show the efficiency of XNorMaj operations on application specific integrated circuits (ASIC) implementations. 

\bibliographystyle{IEEEtran}
% argument is your BibTeX string definitions and bibliography database(s)
\bibliography{bib}

% Generated by IEEEtran.bst, version: 1.14 (2015/08/26)
\begin{thebibliography}{1}
\providecommand{\url}[1]{#1}
\csname url@samestyle\endcsname
\providecommand{\newblock}{\relax}
\providecommand{\bibinfo}[2]{#2}
\providecommand{\BIBentrySTDinterwordspacing}{\spaceskip=0pt\relax}
\providecommand{\BIBentryALTinterwordstretchfactor}{4}
\providecommand{\BIBentryALTinterwordspacing}{\spaceskip=\fontdimen2\font plus
\BIBentryALTinterwordstretchfactor\fontdimen3\font minus
  \fontdimen4\font\relax}
\providecommand{\BIBforeignlanguage}[2]{{%
\expandafter\ifx\csname l@#1\endcsname\relax
\typeout{** WARNING: IEEEtran.bst: No hyphenation pattern has been}%
\typeout{** loaded for the language `#1'. Using the pattern for}%
\typeout{** the default language instead.}%
\else
\language=\csname l@#1\endcsname
\fi
#2}}
\providecommand{\BIBdecl}{\relax}
\BIBdecl

\bibitem{DBLP:journals/corr/CourbariauxB16}
M.~Courbariaux and Y.~Bengio, ``{BinaryNet: Training Deep Neural Networks with
  Weights and Activations Constrained to +1 or -1},'' \emph{CoRR}, vol.
  abs/1602.02830, 2016.

\bibitem{ramin:JHAndersonALMIdea}
J.~H. Kim, J.~Lee, and J.~Anderson, ``{FPGA} architecture enhancement for
  efficient {BNN} implementation,'' in \emph{Int. Conf. on Field-Programmable
  Technology}, 2018, pp. 1--8.

\bibitem{DBLP:conf/fccm/WangDCC19}
E.~Wang, J.~J. Davis, P.~Y.~K. Cheung, and G.~A. Constantinides, ``{LUTNet}:
  Rethinking inference in {FPGA} soft logic,'' in \emph{{IEEE} Annual Int.
  Symp. on Field-Programmable Custom Computing Machines {FCCM}}, 2019.

\bibitem{DBLP:conf/fpga/UmurogluFGBLJV17}
Y.~Umuroglu, N.~J. Fraser, G.~Gambardella, M.~Blott, P.~H.~W. Leong, M.~Jahre,
  and K.~A. Vissers, ``{FINN:} {A} framework for fast, scalable binarized
  neural network inference,'' in \emph{Proc. Int. Symp. on Field-Programmable
  Gate Arrays}, 2017, pp. 65--74.

\bibitem{DBLP:conf/icml/IoffeS15}
S.~Ioffe and C.~Szegedy, ``Batch normalization: Accelerating deep network
  training by reducing internal covariate shift,'' in \emph{Proc. of the 32nd
  Int. Conf. on Machine Learning, {ICML}}, 2015, pp. 448--456.

\bibitem{DBLP:journals/ijon/LiangYLLW18}
S.~Liang, S.~Yin, L.~Liu, W.~Luk, and S.~Wei, ``{FP-BNN:} binarized neural
  network on {FPGA},'' \emph{Neurocomputing}, vol. 275, pp. 1072--1086, 2018.

\bibitem{DBLP:conf/nips/HubaraCSEB16}
I.~Hubara, M.~Courbariaux, D.~Soudry, R.~El{-}Yaniv, and Y.~Bengio, ``Binarized
  neural networks,'' in \emph{Advances in Neural Information Processing Systems
  29: Conf. on Neural Information Processing Systems}, 2016.

\end{thebibliography}
%
% <OR> manually copy in the resultant .bbl file
% set second argument of \begin to the number of references
% (used to reserve space for the reference number labels box)

%\begin{thebibliography}{1}
%\end{thebibliography}

% that's all folks
\end{document}